\documentclass[notitlepage,aps,prx,twoside, superscriptaddress,reprint,showkeys,nofootinbib, onecolumn]{revtex4-1}

\usepackage{amsmath,amsfonts,amssymb,amsthm,amsbsy,mathtools, bbold}
\usepackage{graphicx}
\usepackage{gensymb}
\usepackage[all]{xy}
\usepackage{epstopdf}
\usepackage{bm}

\usepackage{color}

\begin{document}

\title{Indefinite causal structures for continuous-variable systems}

\author{Flaminia Giacomini}
\author{Esteban Castro-Ruiz}
\author{\v{C}aslav Brukner}
\affiliation{Vienna Center for Quantum Science and Technology (VCQ), Faculty of Physics,
University of Vienna, Boltzmanngasse 5, A-1090 Vienna, Austria}
\affiliation{Institute of Quantum Optics and Quantum Information (IQOQI),
Austrian Academy of Sciences, Boltzmanngasse 3, A-1090 Vienna, Austria}

\begin{abstract}
In all our well-established theories, it is assumed that events are embedded in a global causal structure such that, for every pair of events, the causal order between them is always fixed. However, the possible interplay between quantum mechanics and general relativity may require a revision of this paradigm. The process matrix framework keeps the validity of quantum physics locally but does not assume the existence of a global causal order. It allows to describe causal structures corresponding to a quantum superposition of `A is before B' and `B is before A'.  So far, the framework has been developed only for finite-dimensional systems, and a straightforward generalization to infinite dimensions leads to singularities. Such generalization is necessary for continuous-variable systems and is a prerequisite for quantum fields on indefinite causal structures. Here we develop the process-matrix framework for continuous-variable systems. We encounter and solve the problems of singularities. Moreover, we study an example of a process in infinite dimensions, and we derive correlations exhibiting intereference due to processes in which A is before B and B is before A.
\end{abstract}

\maketitle

\section{Introduction}

The notion of causality is deeply rooted in our understanding of nature. In ordinary situations with a fixed spacetime background we can always say that the cause belongs to the past light cone of the effect, and the effect to the future light cone of the cause. This familiar idea might be untenable at regimes at which the quantum mechanical properties of the systems under study are of comparable relevance to their gravitational properties \cite{isham1993canonical, hartle1993spacetime, butterfield2001spacetime}, for instance if the metric tensor, and thus the causal relations, are subject to quantum fluctuations.

The crucial role played by probability in quantum mechanics on the one hand, and the dynamical causal structure of general relativity on the other hand, led to the conjecture that a theory unifying general relativity and quantum mechanics should be a probabilistic theory on a dynamical causal structure \cite{hardy2005probability}. Adopting an operational point of view, we can ask what the measurable consequences of an indefinite causal structure would be. The process matrix framework \cite{oreshkov2012quantum} is a possible way to address this question, and exploits techniques typical of quantum information to deal with the problem. The framework retains the validity of ordinary quantum mechanics at a local level, i.e. in local laboratories where quantum operations are performed, but makes no assumptions on the global causal structure outside the laboratories. Interestingly, the framework allows for processes that are more general than those allowed by the standard (causal) quantum formalism. In particular, they include situations in which the direction of the signaling, and thus causality in the operational sense, is not fixed. Nonetheless, logical paradoxes, such as signaling to the past, are ruled out by consistency conditions. 

We call a process matrix causally ordered if it allows for signalling only in one fixed direction between the parties. A (bipartite) process matrix is {\it causally separable} if it can be decomposed as a convex combination of causally ordered processes. An example of a {\it causally nonseparable} process is the `quantum switch' \cite{oreshkov2012quantum, oreshkov2015causal}. This is a quantum system with an auxiliary degree of freedom which can coherently control the order in which operations are applied. The quantum switch provides quantum computational advantages with respect to quantum circuits with fixed gate order \cite{chiribella2013quantum, chiribella2012perfect, araujo2014computational} and has recently been implemented with linear optics \cite{procopio2014experimental}.

In their original formulation, process matrices were only defined for finite dimensional Hilbert spaces \cite{oreshkov2012quantum, araujo2014computational, araujo2015witnessing, oreshkov2015causal}. Despite providing an arena for the experimental verification of systems like the quantum switch, finite-dimensional systems are too restrictive for the purpose of studying indefinite causality. The generalization of the formalism to continuous variables broadens the class of systems which can be described with the formalism. In particular Gaussian quantum optics, used to describe some cases of continuous-variable quantum systems, has a very important role in quantum information processing \cite{weedbrook2012gaussian}. The generalization proposed here can be straightforwardly used to devise new experiments. As an example of such applications, we propose an infinite-dimensional version of the quantum switch.

In addition, quantum fluctuations of the metric and of the causal structures are expected at high energies, where both quantum and gravitational effects become relevant. At these regimes a description in terms of quantum fields is required. The generalization proposed here is a necessary step towards this goal and paves the way for a more thorough study of quantum fields on indefinite causal structures. With this in mind, it is worth noting that a proper treatment of quantum fields requires to solve problems related to the localisation of the local laboratories and the tensor product structure of the Hilbert spaces. The study of this problem is beyond the scope of this work and is likely to require the framework of algebraic quantum field theory \cite{haag1964algebraic, haag1992local}.

Contrary to the finite-dimensional case, in this work we face difficulties related to singularities. These singularities arise from the straightforward generalization of the approach used in finite dimensions when the dimensions of the Hilbert space tend to infinity. We solve this problem by using a phase space representation of the process matrices in terms of Wigner functions. We also show that the notion of causal nonseparability is maintained in infinite dimensions and we provide an argument for the causal nonseparability of the quantum switch. Specifically, we show that it exhibits interference due to the superposition of the order in which the operations are applied.

\section{The W-matrix formalism}
In this section we give a brief introduction to the W-matrix formalism in finite-dimensional Hilbert spaces, following the first formulation given in \cite{oreshkov2012quantum}. Here we restrict the discussion to a two-party scenario, but the formalism is valid for an arbitrary number or local observers. Let us consider the two observers A and B, situated in separate local laboratories. We assume that standard quantum mechanics is valid in each local laboratory. However, we make no assumptions on the global causal structure outside the laboratories. This means that each observer is free to perform local quantum operations on a physical system in a finite-dimensional Hilbert space. More specifically, the local operations performed in the laboratories are completely positive (CP) maps $\mathcal{M}_i^{A}: \mathcal{L}(\mathcal{H}^{A_1}) \rightarrow \mathcal{L}(\mathcal{H}^{A_2})$ and $\mathcal{M}_j^{B}: \mathcal{L}(\mathcal{H}^{B_1}) \rightarrow \mathcal{L}(\mathcal{H}^{B_2})$, where $\mathcal{L}(\mathcal{H})$ denotes linear operators acting on the finite-dimensional Hilbert space $\mathcal{H}$, and where $\mathcal{H}^{A_1},\, \mathcal{H}^{A_2}$ and $ \mathcal{H}^{B_1}, \, \mathcal{H}^{B_2}$ are respectively the input and output Hilbert spaces of A and B. It is convenient to use the Choi-Jamio{\l}kowski (CJ) isomorphism \cite{jamiolkowski1972linear, choi2000completely}, which associates an operator in the tensor product of two given Hilbert spaces to a map between the two. We write the CJ-equivalent of the local operations as $M_i^{X}= (\mathbb{1}\otimes \mathcal{M}_i^{X})\left| \Phi^+ \right> \left< \Phi^+ \right|$ on $\mathcal{H}^{X_1}\otimes \mathcal{H}^{X_2}$, $X=A,\, B$, where $\left| \Phi^+ \right>= \sum_i \left| i \right>_{X_1} \left| i \right>_{X_1}$ is the maximally entagled state in the input Hilbert space. Given the set of CP maps $\left\lbrace \mathcal{M}_i^{X} \right\rbrace_{i=1}^{n}$ corresponding to all possible $n$ local outcomes, the sum $\sum_i \mathcal{M}_i^X$ is also trace preserving (TP). Physically this means that an outcome always occurs in an experiment. Using the Choi-Jamio{\l}kowski isomorphism, we can write this condition (CPTP condition) as $\sum_i\text{Tr}_{X_2}M_i^X= \mathbb{1}_{X_1}$.

Given the set of CP maps accounting for all possible local operations, we can ask which are the most general correlations between the outcomes of the two observers. The most general way to linearly map local quantum operations to probability distributions can be written as $p(\mathcal{M}_i^A,\, \mathcal{M}_j^B)= \text{Tr}\left[ W (M_i^A \otimes M_j^B) \right]$, where we introduce the process matrix $W \in \mathcal{L}\left( \mathcal{H}^{A_1}\otimes \mathcal{H}^{A_2} \otimes \mathcal{H}^{B_1} \otimes \mathcal{H}^{B_2}\right)$, a positive linear operator $W \geq 0$. The non-negativity of the probabilities (including the case when the two parties share entanglement) is ensured by the positivity of the W-matrix. Moreover, we require that probabilities are normalised, i.e. $\sum_{ij}p(\mathcal{M}_i^A,\, \mathcal{M}_j^B)=1$.

In \cite{araujo2015witnessing} it was shown that the characterization of the W-matrix in the two-party scenario and finite-dimensional Hilbert spaces can be given as
\begin{align} \label{eq:Characterization}
	& W \geq 0, \\
	& \text{Tr}W =d_{A_2} d_{B_2}, \qquad d_X= \text{dim}(\mathcal{H}_{X}), \nonumber\\
	& _{B_1 B_2} W = _{A_2 B_1 B_2} W, \nonumber\\
	& _{A_1 A_2} W = _{B_2 A_1 A_2} W,\nonumber\\
	& W= _{A_2} W + _{B_2} W - _{A_2 B_2} W, \nonumber
\end{align}
where $_{X} W= \frac{\mathbb{1}_X}{d_X} \otimes \text{Tr}_X W$.
This means that not all the subspaces of the space of process matrices are allowed, because they give rise to non-normalized probabilities. In \cite{oreshkov2012quantum} it is shown that these terms can be interpreted as logical paradoxes. As an example, let us assume a one-party scenario in which the input and output Hilbert spaces are two-dimensional and a basis is provided by the two states $\left|0\right>$ and $\left|1\right>$.  Let the W-matrix be an identity channel from the observer's output to the observer's input. Then if the observer applies a local operation which flips the qubit, we get the paradox $\left|0\right>=\left|1\right>$. This paradox is of the type of the `grandfather paradox', in which an agent goes back in time and kills his grandfather. This situations are automatically ruled out in the W-matrix formalism by the conditions \eqref{eq:Characterization}. On the other hand, the requirements \eqref{eq:Characterization} together with the local CPTP maps give rise to correlations which are more general than those of standard quantum mechanics.

In the formulation in finite-dimensional Hilbert spaces the characterization of the process matrix heavily relies on the dimension of the Hilbert spaces of the observers, so that taking the representation of W and letting the dimensions tend to infinity would lead to singularities. Therefore a straightforward generalization to infinite dimensions is not possible. An alternative formulation, suitable for infinite-dimensional Hilbert spaces, is given in terms of Wigner functions, which provide an equivalent description to the usual operator representation. We will see that the requirement that W gives rise to consistent probabilities restricts the possible Wigner representations, and provides an equivalent characterization of the process matrix to the finite-dimensional case.

\section{Extension to infinite dimensions}
The extension of the W-matrix formalism to continuous variables presents some novel features in contrast to the original framework in finite-dimensional Hilbert spaces. These features are analogous to those encountered in the infinite-dimensional limit of ordinary quantum mechanics of finite-dimensional systems \cite{peres2006quantum}, and mainly concern the boundedness of the operators representing a quantum state.

We consider two local observers, $A$ and $B$, each provided with a local laboratory and free to perform local operations on a quantum system. In infinite dimensions we have to restrict the domain $\mathcal{L}(\mathcal{H})$ of linear operators on the Hilbert space $\mathcal{H}$ to bounded linear operators on $\mathcal{H}$. We call this space $\mathcal{B}(\mathcal{H})$. The maps describing the local operations in A and B are represented by completely positive (CP) maps $\mathcal{M}_i^{A}: \mathcal{B}(\mathcal{H}_{A_{1}}) \rightarrow \mathcal{B}(\mathcal{H}_{A_{2}})$, $\mathcal{M}_j^{B}: \mathcal{B}(\mathcal{H}_{B_{1}}) \rightarrow \mathcal{B}(\mathcal{H}_{B_{2}})$, where $\mathcal{H}_{X_{1}}, \, \mathcal{H}_{X_{2}}$, $X=A,\,B$, are the (infinite-dimensional) input and output Hilbert spaces of each laboratory. Each map $\mathcal{M}_i^{X}$ describes transformations of a state $\rho$ with outcome $i$ and output state $\mathcal{M}_i^{X}(\rho)$. A convenient way of representing CP maps is through the Choi-Jamiolkowski (CJ) isomorphism (see \cite{jamiolkowski1972linear, choi2000completely} for the original definition in finite dimensions,  \cite{holevo2011entropy} for the extension to infinite dimensions), which associates an operator $M_i^X$ to a CP map $\mathcal{M}_i^X$ through
	$M_i^{X}= \left( \mathbb{1} \otimes \mathcal{M}_i^{X} \right) \left| \Phi^+ \right> \left< \Phi^+ \right|$.
Here $\left| \Phi^+ \right>= \int dx \left| xx \right>_{X_{1}}$ is the non-normalized maximally entangled state in $\mathcal{H}_{X_{1}} \otimes \mathcal{H}_{X_{1}}$ and $\mathbb{1}$ is the identity operator. Since the probability of obtaining an outcome is unity, the sum over all possible $\mathcal{M}_i^X$ is a completely positive trace-preserving (CPTP) map. This condition, which we refer to as CPTP condition, is expressed in terms of the CJ equivalent $M^X= \sum_i M_i^X$ as $\operatorname{Tr}_{X_{2}}(M^{X})=\mathbb{1}_{X_{1}}$. 

The process matrix is an operator $W \in \mathcal{B}(\mathcal{H}_{A_{1}} \otimes \mathcal{H}_{A_{2}}\otimes \mathcal{H}_{B_{1}}\otimes \mathcal{H}_{B_{2}})$ such that  $W \geq 0$ and the probability of two measurement outcomes $i$ and $j$ is
\begin{equation}
	p(\mathcal{M}_i^{A},\, \mathcal{M}_j^{B}) = \operatorname{Tr} \left[ W (M_i^{A} \otimes M_j^{B}) \right].
\end{equation}
The probability should satisfy $0 \leq p(\mathcal{M}_i^{A},\, \mathcal{M}_j^{B}) \leq 1$. In particular, the condition $\sum_{ij} p(\mathcal{M}_i^{A},\, \mathcal{M}_j^{B}) = 1$ implies that $\operatorname{Tr}\left[W (M^{A}\otimes M^{B})\right]=1$ for every pair of CPTP maps $\mathcal{M}^{A},\, \mathcal{M}^{B}$. From now on we will only consider the CJ representation of the CP maps. 

\subsection{Characterization of the one-party scenario}
The one party scenario can be obtained from the two parties when the Hilbert spaces of one observer are one-dimensional.
The Wigner equivalent of a CPTP map $M$ (we omit here the index relative to the observer) and of a process matrix $W$ is a function of four variables on the phase space, namely $M(\boldsymbol{\xi}_{1}, \boldsymbol{\xi}_{2})$ and $W(\boldsymbol{\xi}_{1}, \boldsymbol{\xi}_{2})$. Here the subscripts $1$ and $2$ refer respectively to the input and output Hilbert space and the quantity $\boldsymbol{\xi}_i$ corresponds to the point in the phase space $\boldsymbol{\xi}_i=(x_i, p_i)$. In terms of Wigner functions, the CPTP condition becomes
	$\frac{1}{2\pi} \int d\boldsymbol{\xi}_{2} M (\boldsymbol{\xi}_{1}, \boldsymbol{\xi}_{2})=1$.
By computing the Fourier transform $\tilde{M}(\boldsymbol{\eta}_{1}, \boldsymbol{\eta}_{2})= \frac{1}{(2\pi)^2}\int d\boldsymbol{\xi}_{1} d\boldsymbol{\xi}_{2} M(\boldsymbol{\xi}_{1}, \boldsymbol{\xi}_{2}) e^{-i \boldsymbol{\xi}_{1} \cdot \boldsymbol{\eta}_{1}}e^{-i \boldsymbol{\xi}_{2} \cdot \boldsymbol{\eta}_{2}}$, with $\boldsymbol{\eta}_i= (\kappa_i, \omega_i)$ the previous condition reads
	\begin{equation} \label{eq:CPTPcondition}
		\tilde{M} (\boldsymbol{\eta}_{1},\boldsymbol{0})= 2\pi \delta(\boldsymbol{\eta}_{1}),
	\end{equation}
where $\delta(\boldsymbol{\eta}_1)= \delta(\kappa_1)\delta(\omega_1)$ and $\delta$ is the Dirac delta function.

We use the CPTP condition \eqref{eq:CPTPcondition} to characterize the $W$-matrix. In terms of the Wigner representation the normalization of probability $Tr(W M^A)=1$ is
\begin{equation} \label{eq:norm_oneparty}
	\frac{1}{(2\pi)^2} \int d \boldsymbol{\eta}_{1} d \boldsymbol{\eta}_{2}  \tilde{W} (\boldsymbol{\eta}_{1}, \boldsymbol{\eta}_{2})\tilde{M} (\boldsymbol{\eta}_{1}, \boldsymbol{\eta}_{2})=1.
\end{equation}	
For each $\tilde{M}(\boldsymbol{\eta}_{1}, \boldsymbol{\eta}_{2})$ we identify a small interval $S_{2}(\tilde{M}) \in \mathbb{R}^2$ around $\boldsymbol{\eta}_{2} = \boldsymbol{0}$ where we can approximate  $\tilde{M} (\boldsymbol{\eta}_{1}, \boldsymbol{\eta}_{2})$ with $\tilde{M} (\boldsymbol{\eta}_{1}, \boldsymbol{0})$. We assume that the  function $\tilde{M}$ has a well-defined limit at $\boldsymbol{\eta}_2=0$. For all possible $\tilde{M} (\boldsymbol{\eta}_{1}, \boldsymbol{\eta}_{2})$ we choose the smallest interval $S_{2}=\min_{\tilde{M}} S_{2}(\tilde{M})$. We set
	$S_{2} \equiv \left[ -\frac{\epsilon}{2},\,\frac{\epsilon}{2}\right]\times \left[ -\frac{\delta}{2},\,\frac{\delta}{2}\right]$.
We now split our integral in two parts: in the first one the output variables are integrated over $S_{2}$; in the second one the integration is performed on $\mathbb{R}^2\setminus S_{2}$. By using equation \eqref{eq:CPTPcondition} in the integral on $S_{2}$, equation \eqref{eq:norm_oneparty} reads
\begin{equation} \label{eq:splitoneparty}
	1= \frac{\epsilon \delta}{2\pi} \tilde{W} (\boldsymbol{0}, \boldsymbol{0})  + \left<\tilde{W}\tilde{M}\right>_{\mathbb{R}^2, \mathbb{R}^2\setminus S_{2}},
\end{equation}
where $\left< f \right>_{R_i, R_j}= \frac{1}{(2\pi)^2}\int_{R_i} d \boldsymbol{\eta}_{1} \int_{R_j} d \boldsymbol{\eta}_{2} f(\boldsymbol{\eta}_{1}, \boldsymbol{\eta}_{2})$. Note that, in order to satisfy equation \eqref{eq:splitoneparty}, $\tilde{W}(\boldsymbol{\eta}_1,\boldsymbol{0})$ can not diverge faster than $1/\epsilon \delta$. This implies that for all possible $\tilde{M} (\boldsymbol{\eta}_{1}, \boldsymbol{\eta}_{2})$, restricted to the domain $\mathbb{R}^2 \times (\mathbb{R}^2 \setminus S_2)$, the second term in the sum is always equal to the same constant. This can only happen if the second term in the sum in equation \eqref{eq:splitoneparty} vanishes, so we conclude that 
	$\tilde{W} (\boldsymbol{\eta}_{1}, \boldsymbol{\eta}_{2})=0$ when $\boldsymbol{\eta}_{2} \notin S_{2}$ and $\tilde{W} (\boldsymbol{0}, \boldsymbol{\eta}_{2}) = \frac{2\pi}{\epsilon \delta}$ when $\boldsymbol{\eta}_{2} \in S_{2}$.
We now send $\epsilon$ and $\delta$ to zero. In the limit we find
\begin{equation} \label{eq:Woneparty}
	\tilde{W} (\boldsymbol{\eta}_{1}, \boldsymbol{\eta}_{2})= 2 \pi w(\boldsymbol{\eta}_{1}) \delta(\boldsymbol{\eta}_{2}),
\end{equation}
where $w(\boldsymbol{\eta}_1)$ is a function to be determined.

We now ask which conditions $w(\boldsymbol{\eta}_{1})$ should satisfy in order for the probability to be normalized. If we substitute the result \eqref{eq:Woneparty} in the condition for the normalization of the probability \eqref{eq:norm_oneparty} we see that
	$1=	\frac{1}{2\pi} \int d \boldsymbol{\eta}_{1} w(\boldsymbol{\eta}_{1}) \tilde{M} (\boldsymbol{\eta}_{1},\boldsymbol{0})= w(\boldsymbol{0})$.
Moreover, we can write the complete expression for the Wigner function as
	$W (\boldsymbol{\xi}_{1}, \boldsymbol{\xi}_{2}) 
	=\frac{1}{2\pi} \int d \boldsymbol{\eta}_{1} e^{i\boldsymbol{\xi}_{1} \cdot \boldsymbol{\eta}_{1}}w (\boldsymbol{\eta}_{1})$.
The Wigner equivalent of the $W$-matrix does not depend on the variables of the second Hilbert space. In the operator representation this result is equivalent to having the identity in the second Hilbert space. This is compatible with the finite-dimensional case shown in \cite{oreshkov2012quantum}. Moreover, given $W = W_1 \otimes \mathbb{1}_2$, computing the partial trace on the first system leads to
	$Tr_{1} W_1 =\frac{1}{(2\pi)^2} \int d \boldsymbol{\xi}_{1} d \boldsymbol{\eta}_{1} e^{i\boldsymbol{\xi}_{1} \cdot \boldsymbol{\eta}_{1}} w (\boldsymbol{\xi}_{1})= w(\boldsymbol{0})=1$.
This means that in $\mathcal{H}_1$ the $W$-matrix is a state with unit trace. Therefore, the most general form of the total $W$ for the one-party case is $W= \rho \otimes \mathbb{1}$, consistent with the finite-dimensional case.

\subsection{Characterization of the two-party scenario}
In the bipartite case the Wigner equivalent of the $W$-matrix is a function of eight variables in the phase space $W(\boldsymbol{\xi}_{A_{1}}, \boldsymbol{\xi}_{A_{2}}, \boldsymbol{\xi}_{B_{1}}, \boldsymbol{\xi}_{B_{2}})$, where the notation is consistent with the previous case.
The probability normalization in terms of the Fourier transform of the Wigner equivalents of the operators is
\begin{align} \label{eq:normprobAB}
	1=&\frac{1}{(2\pi)^4} \int d \boldsymbol{\eta}_{A_{1}} d \boldsymbol{\eta}_{A_{2}} d \boldsymbol{\eta}_{B_{1}} d \boldsymbol{\eta}_{B_{2}}\tilde{W}(\boldsymbol{\eta}_{A_{1}}, \boldsymbol{\eta}_{A_{2}}, \boldsymbol{\eta}_{B_{1}}, \boldsymbol{\eta}_{B_{2}})\nonumber \\ 
	& \times\tilde{M}^{A}(\boldsymbol{\eta}_{A_{1}}, \boldsymbol{\eta}_{A_{2}}) \tilde{M}^{B}( \boldsymbol{\eta}_{B_{1}}, \boldsymbol{\eta}_{B_{2}}),
\end{align}
where the CPTP condition for $\tilde{M}^A$ and $\tilde{M}^B$ is described by equation \eqref{eq:CPTPcondition}.
Consider now a specific local operation for one of the two parties, say Alice, given by
$\tilde{M}^{A}(\boldsymbol{\eta}_{A_{1}}, \boldsymbol{\eta}_{A_{2}}) = 2\pi \delta(\boldsymbol{\eta}_{A_{1}}) \chi(R_{A_{2}})$, where $\chi(R_{A_{2}})$ is the characteristic function over the set $R_{A_{2}}$, $\chi(R_{A_{2}})= 1$ when $\boldsymbol{\eta}_{A_{2}} \in R_{A_{2}}$, $\chi(R_{A_{2}})= 0$ otherwise. $R_{A_{2}}$ is a two-dimensional set defined as $R_{A_{2}} = \left[ -\frac{1}{2\alpha_1},\frac{1}{2\alpha_1} \right]\times \left[ -\frac{1}{2\alpha_2},\frac{1}{2\alpha_2} \right]$ and $\alpha_1, \, \alpha_2$ are two arbitrary positive numbers. This choice of the measurement satisfies the CPTP condition for all $\alpha_1,\, \alpha_2$. By inserting this in equation \eqref{eq:normprobAB} we obtain
\begin{align}
	1=&\frac{1}{(2\pi)^3}\frac{\alpha_1 \alpha_2}{\alpha_1 \alpha_2} \int d \boldsymbol{\eta}_{A_{2}} d \boldsymbol{\eta}_{B_{1}} d \boldsymbol{\eta}_{B_{2}} \tilde{W}(\boldsymbol{0}, \boldsymbol{\eta}_{A_{2}}, \boldsymbol{\eta}_{B_{1}}, \boldsymbol{\eta}_{B_{2}})\nonumber \\
	&\times\chi(R_{A_{2}}) \tilde{M}^{B}(\boldsymbol{\eta}_{B_{1}}, \boldsymbol{\eta}_{B_{2}}) \nonumber
\end{align}
If we now let $\alpha_1, \alpha_2$ be very large, but still finite, we can approximate $\alpha_1 \alpha_2 \chi(R_{A_{2}})$ with the product of two delta functions, so that we can perform the integration in $\boldsymbol{\eta}_{A_{2}}$ by evaluating the $W$-matrix in the origin. Therefore, the condition to impose on the total $W$ to have an integral converging to a constant (one) is
	$\tilde{W}(\boldsymbol{0},\, \boldsymbol{\eta}_{A_{2}}, \boldsymbol{\eta}_{B_{1}}, \boldsymbol{\eta}_{B_{2}})= 2\pi\alpha_1 \alpha_2 \tilde{W}_{B}(\boldsymbol{\eta}_{B_{1}}, \boldsymbol{\eta}_{B_{2}})$
 whenever $\boldsymbol{\eta}_{A_{2}} \in R_{A_{2}}$ and $W=0$ otherwise. $\tilde{W}_{B}(\boldsymbol{\eta}_{B_{1}}, \boldsymbol{\eta}_{B_{2}})$ is the reduced $W$ of the observer B. As a consequence, in the limit $\alpha_1,\,\alpha_2 \rightarrow \infty$ we obtain
$1=\frac{1}{(2\pi)^2} \int d \boldsymbol{\eta}_{B_{1}} d \boldsymbol{\eta}_{B_{2}}  \tilde{W}_{B}(\boldsymbol{\eta}_{B_{1}}, \boldsymbol{\eta}_{B_{2}})\tilde{M}^{B}( \boldsymbol{\eta}_{B_{1}}, \boldsymbol{\eta}_{B_{2}})$.
The previous equation describes exactly the one-party case, so we can apply the result \eqref{eq:Woneparty} and write
\begin{equation} \label{eq:middlecondW_A}
	\tilde{W}(\boldsymbol{0},\, \boldsymbol{\eta}_{A_{2}}, \boldsymbol{\eta}_{B_{1}}, \boldsymbol{\eta}_{B_{2}})= (2 \pi)^2 \tilde{w}_{B_{1}}(\boldsymbol{\eta}_{B_{1}})\delta(\boldsymbol{\eta}_{B_{2}})\delta(\boldsymbol{\eta}_{A_{2}}).
\end{equation}
This decomposition of $\tilde{W}$ is correct only if $\boldsymbol{\eta}_{A_{2}}$ is arbitrarily close to the origin. If we now repeat the same procedure by swapping the measurements of Alice and Bob we find an analogous condition
\begin{equation} \label{eq:middlecondW_B}
	\tilde{W}(\boldsymbol{\eta}_{A_1},\, \boldsymbol{\eta}_{A_{2}}, \boldsymbol{0}, \boldsymbol{\eta}_{B_{2}})= (2\pi)^2\tilde{w}_{A_{1}}(\boldsymbol{\eta}_{A_{1}})\delta(\boldsymbol{\eta}_{A_{2}})\delta(\boldsymbol{\eta}_{B_{2}}),
\end{equation}
which holds when $\boldsymbol{\eta}_{B_{2}}$ is arbitrarily close to the origin. 

We now go back to the equation \eqref{eq:normprobAB} for the normalization of probability. Similarly to the one-party case, we define two intervals
$S_{A_{2}} = \left[ -\frac{\epsilon_A}{2},\frac{\epsilon_A}{2} \right]\times \left[ -\frac{\delta_A}{2},\frac{\delta_A}{2} \right] \in \mathbb{R}^2$ and
	$S_{B_{2}} = \left[ -\frac{\epsilon_B}{2},\frac{\epsilon_B}{2} \right]\times \left[ -\frac{\delta_B}{2},\frac{\delta_B}{2} \right] \in \mathbb{R}^2$, where we can approximate the functions $\tilde{M}^{A}$ and $\tilde{M}^{B}$ with their values in respectively $\boldsymbol{\eta}_{A_{2}} = \boldsymbol{0}$ and $\boldsymbol{\eta}_{B_{2}} = \boldsymbol{0}$. We can now split the probability condition in four parts, writing the integrals over $A_2$ and $B_2$ as the sum of an integral over $S_{A_2}$ and $S_{B_2}$ and on the rest of the integration region $\bar{S}_{A_{2}}$ and $\bar{S}_{B_{2}}$. Using the CPTP condition for the local operations we find
\begin{equation}
	1=P_{S_{A_{2}},S_{B_{2}}}+ P_{S_{A_{2}}, \bar{S}_{B_{2}}} + P_{\bar{S}_{A_{2}}, S_{B_{2}}} + P_{\bar{S}_{A_{2}},\bar{S}_{B_{2}}}
\end{equation}
where
\begin{align*}
	&P_{S_{A_{2}},S_{B_{2}}}= const,\\
	&P_{S_{A_{2}}, \bar{S}_{B_{2}}}= k_A \int d \boldsymbol{\eta}_{B_{1}} \tilde{w}_{B_{1}}(\boldsymbol{\eta}_{B_{1}})\int_{\mathbb{R}^2 \setminus S_{B_{2}}}d\boldsymbol{\eta}_{B_{2}}\delta(\boldsymbol{\eta}_{B_{2}}),\\
	&P_{\bar{S}_{A_{2}}, S_{B_{2}}}= k_B \int d \boldsymbol{\eta}_{A_{1}} \tilde{w}_{A_{1}}(\boldsymbol{\eta}_{A_{1}})\int_{\mathbb{R}^2 \setminus S_{A_{2}}}d\boldsymbol{\eta}_{A_{2}}\delta(\boldsymbol{\eta}_{A_{2}}),\\
	&P_{\bar{S}_{A_{2}},\bar{S}_{B_{2}}}= \left< \tilde{W}\tilde{M}^{A} \tilde{M}^{B}\right>_{\mathbb{R}^2, \mathbb{R}^2, \mathbb{R}^2\setminus S_{A_{2}}, \mathbb{R}^2 \setminus S_{B_{2}}}.
\end{align*}
Here, $k_A,\ k_B$ are constants and the notation for the last term is analogous to the one used in the one-party case. $P_{S_{A_{2}}, \bar{S}_{B_{2}}}$ and $P_{\bar{S}_{A_{2}}, S_{B_{2}}}$ are identically zero because the delta functions vanish in the interval.
Since the integral is equal to the same constant for all local operations we conclude that the fourth term is zero in the interval considered. For this to be the case, the $W$-function should be zero outside $S_{A_{2}}$ or $S_{B_{2}}$, at least in one of the outputs. Setting $\tilde{W}$ equal to zero in the input would instead lead to the trivial solution $W=0$. By taking the limit when the intervals $S_{A_{2}}, S_{B_{2}}$ reduce to a point, and following an analogous procedure to the one-party case, it is possible to show that the $W$-matrix is a delta function at least in one of the two outputs. Applying the inverse Fourier transform, in the original variables $\boldsymbol{\xi}_i$ the conditions on the $W$ imply that the Wigner equivalent of the process matrix can not depend on both outputs at the same time, i.e. $W(\boldsymbol{\xi}_{A_{1}}, \boldsymbol{\xi}_{A_{2}}, \boldsymbol{\xi}_{B_{1}})$ or $W(\boldsymbol{\xi}_{A_{1}}, \boldsymbol{\xi}_{B_{1}}, \boldsymbol{\xi}_{B_{2}})$. As we have already pointed out in the one-party scenario, this condition is equivalent to having an identity in at least one of the two output Hilbert spaces when $W$ is represented in the space of linear operators on the tensor product of the four Hilbert spaces.

The results for the infinite-dimensional process matrices show that the bipartite $W$ allows for three different situations. The first case consists in a shared state between $A$ and $B$ with no-signaling between the two observers. In the framework of infinite-dimensional W-matrices this is described as $W(\boldsymbol{\xi}_{A_1}, \boldsymbol{\xi}_{B_1})$. The fact that W does not depend on the output variables corresponds to the condition, shown in \cite{oreshkov2012quantum}, $W= \rho_{A_{1} B_{1}} \otimes \mathbb{1}_{A_{2} B_{2}}$. The second and third case describe signaling from one observer to the other. In this case the W-matrix is written as $W(\boldsymbol{\xi}_{A_1}, \boldsymbol{\xi}_{B_1}, \boldsymbol{\xi}_{B_2})$, with correlations at least between $\boldsymbol{\xi}_{A_1}$ and $\boldsymbol{\xi}_{B_2}$, when B signals to A or as $W(\boldsymbol{\xi}_{A_1}, \boldsymbol{\xi}_{A_2}, \boldsymbol{\xi}_{B_1})$, where at least $\boldsymbol{\xi}_{B_1}$ and $\boldsymbol{\xi}_{A_2}$ are correlated, when A signals to B. These two terms are described respectively as $W_{A_{1} A_{2} B_{1}} \otimes \mathbb{1}_{B_{2}}$ and $W_{A_{1} B_{1} B_{2}} \otimes \mathbb{1}_{A_{2}}$ in the finite-dimensional case.

We are interested in processes, which we refer to as \emph{causally nonseparable}, where it is not possible to decompose the $W$-matrix as \cite{araujo2015witnessing, oreshkov2015causal}
\begin{equation} \label{eq:causallyseparable}
	W= \lambda W^{A \prec B} + (1-\lambda)W^{B \prec A},
\end{equation}
where $0 \leq \lambda \leq 1$. If equation \eqref{eq:causallyseparable} holds, the $W$-matrix can always be understood as a classical (convex) mixture of a term which allows signaling from A to B with probability $\lambda$ and a term which allows signaling from B to A with probability $1-\lambda$. The possibility for A and B to share an entangled state with no-signaling correlations is also included in equation \eqref{eq:causallyseparable}.

\section{Quantum switch in infinite dimensions}

\begin{figure}
	\centering
	\includegraphics[scale=0.35]{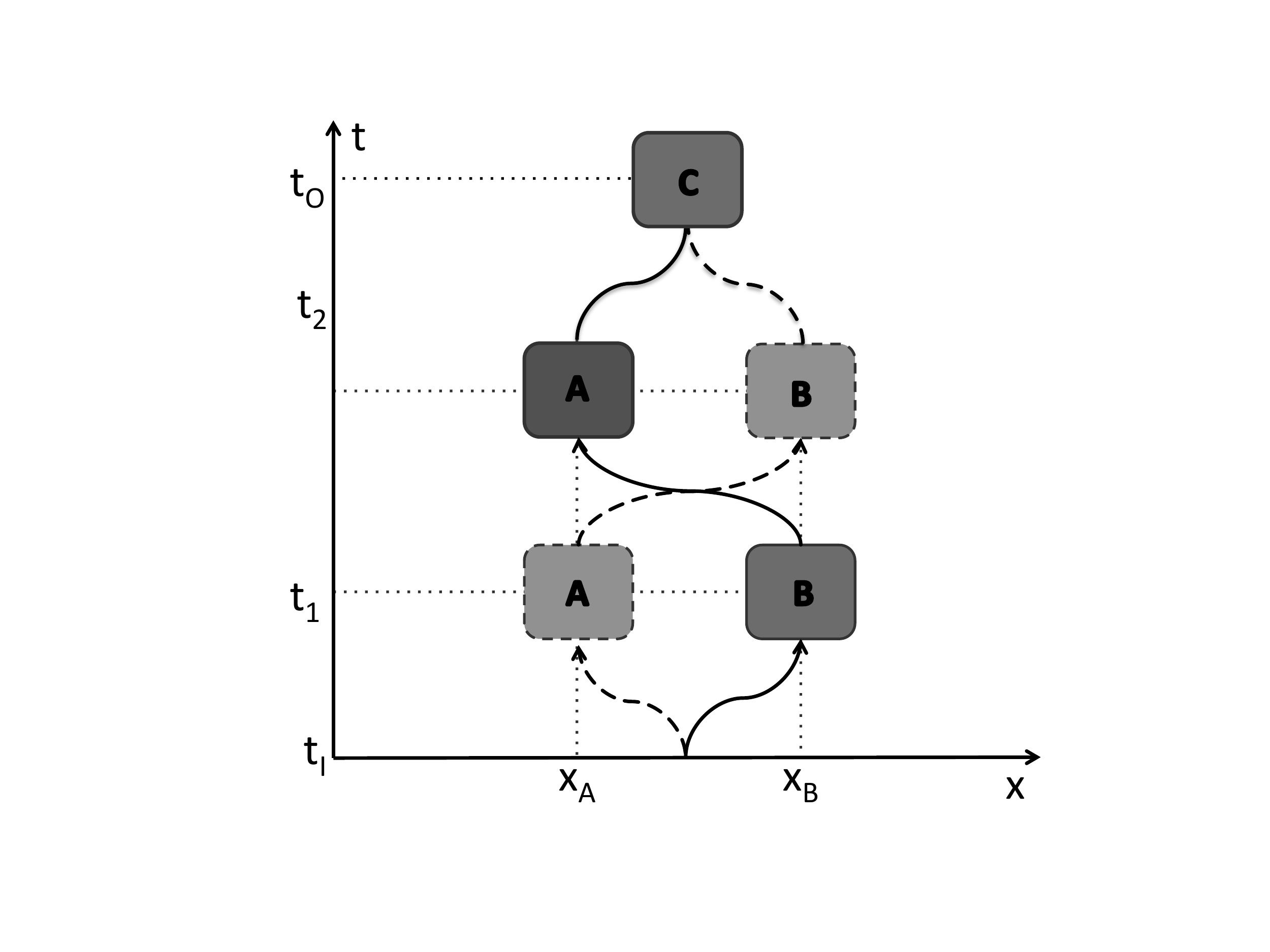}
	\caption{A quantum system is prepared in a state $\left| \psi_I \right>$ at time $t_I$ and is sent in a superposition of two paths. Each path, realized by sending the particle through a fiber (solid and dotted line in the figure), enters the two laboratories A and B in a fixed order and is detected by C at time $t_O$ after exiting the two laboratories. In each local laboratory the state undergoes local quantum operations described as measurement and repreparation. The probability of measurement outcomes shows an interference pattern due to the superposition of two causal orders. The interference can not be reproduced from local operations performed in a fixed causal order.}
	\label{fig:switch}
\end{figure}
A scheme of the quantum switch is provided in Figure \ref{fig:switch}. The switch involves three local observers, which we denote as A, B and C. The observers perform local quantum operations, here chosen to be a measurement followed by a repreparation of a quantum state. Outside the laboratories the system propagates along two ``fibers'' (solid and dotted line in Figure \ref{fig:switch}), which represent the propagation of the quantum system along an additional spatial degree of freedom. A quantum state $\left|\psi_I \right>$ is prepared at time $t_I$ and sent in a superposition of two paths. In one of the paths the particle enters laboratory A at time $t_1$ and laboratory B at time $t_2 > t_1$; in the second path the order of the operations A and B is reversed. After exiting the laboratories A and B the system is detected by the observer C at time $t_O$. Note that in order to preserve the coherence of the process the measurements should not reveal the time.
 
The switch describes a quantum process in which the order of the local operations is in a superposition. In finite dimensions it has been proved that the $W$-matrix which describes the switch is causally nonseparable \cite{araujo2015witnessing}, i.e. it can not be written as $W=\lambda W^{A \prec B \prec C}+(1-\lambda)W^{B \prec A \prec C}$, where C always comes after A and B and $0 \leq\lambda \leq 1$. Here we generalize the switch to infinite dimensions, and provide an alternative proof of its causal nonseparability.

The $W$-matrix is an operator acting on the tensor product of six Hilbert spaces, $W \in \mathcal{B}(\mathcal{H}_{A_1}\otimes \mathcal{H}_{A_2} \otimes \mathcal{H}_{B_1} \otimes \mathcal{H}_{B_2} \otimes \mathcal{H}_{C_1} \otimes \mathcal{H}_{p})$. The first five spaces are infinite-dimensional and $\mathcal{H}_{p}$ is a two-dimensional Hilbert space spanned by the vectors $\left| 0 \right>$ and $\left| 1 \right>$, which label each of the paths (fibers) taken by the particle (see Figure \ref{fig:switch}). The W-matrix of the switch is pure and can be written as
	$W= \left| w \right> \left< w \right|$,
where $\left| w \right>= \int d\bar{r}\, w(\bar{r})\left| \bar{r} \right>$, with $\bar{r}=(r_{A_{1}},\ r_{A_{2}},\ r_{B_{1}}, \ r_{B_{2}}, r_{C_{1}})$. Explicitly,
\begin{equation} \label{eq:Wfunction}
	w(\bar{r})= \frac{1}{\sqrt{2}}\int dr_I \psi_I(r_I) \left( w^{A \prec B \prec C}\left| 0 \right> + w^{B \prec A\prec C} \left| 1 \right>\right).
\end{equation} 
Here, $\psi_I (r_I)$ is a normalized square-integrable function. The variables of the functions $w^{A \prec B \prec C}=w^{A \prec B \prec C}(r_I, \bar{r})$ and $w^{B \prec A \prec C}=w^{B \prec A \prec C}(r_I,\bar{r})$, where the arguments parametrize the propagation along the fiber, are omitted in \eqref{eq:Wfunction} for simplicity. The total state $\left| w \right>$ is a superposition of two terms, decribed by $w^{A \prec B \prec C}$ and $w^{B \prec A \prec C}$, which can be explicitly written as
\begin{align}
	w^{A \prec B \prec C} = &G_{I1}(r_{A_{1}}-r_{I}) G_{12}(r_{B_{1}}-r_{A_{2}})G_{2O}(r_{C_1}-r_{B_{2}}) \label{eq:wABC}\\
	w^{B \prec A \prec C} = &G_{I1}(r_{B_{1}}-r_{I}) G_{12}(r_{A_{1}}-r_{B_{2}})G_{2O}(r_{C_1}-r_{A_{2}}) \label{eq:wBAC}
\end{align}
where $G_{ab}(r_b-r_a)= \left< r_b \right| e^{-\frac{i}{\hbar}\hat{H}(t_b - t_a)}\left| r_a \right>$ is the Green function between $r_a$ and $r_b$ and $\hat{H}$ is the hamiltonian which generates the evolution along the fiber.

Consider now the local operations performed by one of the parties, say A. Suppose that A measures the state in a region $R_i$ of the whole laboratory A. Afterwards, the state is reprepared in $\left| \phi_A \right>$. The Choi-Jamio{\l}kowski equivalent of this local operation in A's laboratory is
	$M_i^A = \int_{R_i} dy_A \left| y_A \right> \left< y_A \right| \otimes  \left| \phi_A \right>\left< \phi_A \right|$. 
The intervals $R_i$ satisfy $R_i \cap R_j= \emptyset$ for $i \neq j$ and $\cup_i R_i= V_{A}$, where $V_A$ is the volume of the local laboratory. The same considerations are valid for the case of B. The observer C detects the state he receives by projecting it over the region $R_k$ of the volume of his laboratory $V_C$ and by recombining the two paths via a measurement on the $\left| \pm \right>= (\left| 0 \right>\pm \left| 1 \right>)/\sqrt{2}$ basis. As a consequence, the local operation performed by C is $M_{k \pm}^C= M_k^C \otimes \left| \pm \right>\left< \pm \right|$, where $M_k^C= \int_{R_k} dy_C \left| y_C \right> \left< y_C \right|$ and it is implied that the output Hilbert space of C is one-dimensional.

The probability of the measurement outcomes is then given by
	$p_{ijk \pm}= p(\mathcal{M}_i^A, \ \mathcal{M}_j^B, \mathcal{M}_{k \pm}^C)= \left< w\right| (M_i^A \otimes M_j^B \otimes M_{k \pm}^C) \left| w \right>$. For simplicity we first consider a density of probability $\Pi_{ijk \pm}=\Pi_{ijk \pm}(r_I, r'_I)$ such that
	$p_{ijk \pm}= \int dr_I dr'_I \psi_I (r_I)  \psi^*_I (r'_I) \Pi_{ijk \pm}(r_I, r'_I)$.
Then we can write
\begin{equation} \label{eq:densityswitch}
	\Pi_{ijk \pm}= \frac{1}{2} \left[ \pi^{A \prec B \prec C}_{ijk \pm} + \pi^{B \prec A \prec C}_{ijk \pm} + 2 \operatorname{Re} \pi^{int}_{ijk \pm}\right],
\end{equation}
where we can express the single terms in the sum by adopting a vector notation with $\left| w^{A\prec B \prec C} \right> = \int d \bar{r}\, w^{A\prec B \prec C} \left| \bar{r} \right>$ and $\left| w^{B\prec A \prec C} \right> = \int d \bar{r}\, w^{B\prec A \prec C} \left| \bar{r} \right>$,
\begin{align}
	&\pi^{A \prec B \prec C}_{ijk \pm}=\frac{1}{2}\left< w^{A\prec B\prec C} \right| M_i^A \otimes M_j^B \otimes M_{k}^C \left| w^{A\prec B \prec C} \right> \nonumber\\
	&\pi^{B \prec A \prec C}_{ijk \pm}=\frac{1}{2}\left< w^{B\prec A \prec C} \right| M_i^A \otimes M_j^B \otimes M_k^C \left| w^{B\prec A \prec C}\right> \nonumber\\
	&\pi^{int}_{ijk \pm}=\pm\frac{1}{2}\left< w^{A\prec B \prec C} \right| M_i^A \otimes M_j^B \otimes M_k^C \left| w^{B\prec A \prec C} \right>.
\end{align}

Assuming $t_1-t_I=t_2-t_1=t_O-t_2=\Delta t$, we can show that $p_{ijk \pm}$ describes a two-way signaling from A to B to C and from B to A to C. Specifically, we show that the two terms $\pi_{ijk \pm}^{A \prec B \prec C}$ and $\pi_{ijk \pm}^{B \prec A \prec C}$ correspond to a process in which the order of the events is fixed. Instead, $\pi^{int}_{ijk \pm}$ is an interference term, due to the superposition of causal orders, describing a two-way signaling between the three observers. In order to show this we can sum over the outputs of the observers and show how the marginals depend on the settings $\phi_A$ of $M_i^A$ and $\phi_B$ of $M_j^B$.

We assume that the states $\psi_I, \phi_A$ and $\phi_B$ are prepared so that the probability of detection in the three local laboratories is almost one. This means that the integration over the volume of any local laboratory (A, B or C) can be extended to an integral over the whole space, since this would amount to adding a negligible term to the sum. Defining $p^{ABC}(ijk\pm | \phi_A, \phi_B)= \int d r_I d r'_I \psi_I (r_I) \psi^*_I (r'_I)\pi_{ijk\pm}^{A \prec B \prec C}$ the integral of the first term in equation \eqref{eq:densityswitch}, we find that $\sum_{jk \pm}p^{ABC}(ijk\pm | \phi_A, \phi_B)= p^{ABC}(i)$, which means that A does not receive information from B and C. Moreover, since $\sum_{ij}p^{ABC}(ijk\pm | \phi_A, \phi_B)= p^{ABC}(k\pm | \phi_B)$, C receives information from B. Finally, the fact that $\sum_{ik\pm}p^{ABC}(ijk\pm | \phi_A, \phi_B)= p^{ABC}(j | \phi_A)$ means that B receives information from A but not from C. Therefore, we conclude that the probability describes a causally ordered process where A signals to B and B signals to C. The situation is symmetric under the exchange of A and B if we consider the integral of the second term in equation \eqref{eq:densityswitch}, $p^{BAC}(ijk\pm | \phi_A, \phi_B)= \int d r_I d r'_I \psi_I (r_I) \psi^*_I (r'_I)\pi_{ijk\pm}^{B \prec A \prec C}$.

A probabilistic mixture of the two terms corresponds to a process with no fixed causal order, but causally separable in the sense previously discussed. In contrast, when the quantum switch is considered an additional interference term appears. The interference corresponds to $\pi^{int}_{ijk \pm}$ in equation \eqref{eq:densityswitch} and it can be shown to be
\begin{align}
	&\pi^{int}_{ijk\pm}= \pm \frac{1}{2}\int_{R_i} dr_{A_1} \int_{R_j} dr_{B_1} \int_{R_k} dr_{C_1} \times \nonumber\\
	&\int dr_{A_2} dr'_{A_2} dr_{B_2} dr'_{B_2} w^{A\prec B \prec C}w^{*B\prec A \prec C}\times \nonumber\\
	&\phi_A(r'_{A_2})\phi^*_A(r_{A_2})\phi_B(r'_{B_2})\phi^*_B(r_{B_2}),
\end{align}
where $w^{A\prec B \prec C}= w^{A\prec B \prec C}(r_I, r_{A_1}, r_{A_2}, r_{B_1}, r_{B_2}, r_{C_1})$ and $w^{B\prec A \prec C}=w^{B\prec A \prec C}(r'_I, r_{A_1}, r'_{A_2}, r_{B_1}, r'_{B_2}, r_{C_1})$ were defined in equations \eqref{eq:wABC} and \eqref{eq:wBAC}. To show that there is two-way signaling, we define $p^{int}(ijk \pm| \phi_A, \phi_B)=\int d r_I d r'_I \psi_I (r_I) \psi^*_I (r'_I)\pi_{ijk\pm}^{int}$ and sum over the outputs of the three observers. We find that $\sum_{ij}p^{int}(ijk \pm| \phi_A, \phi_B)= p^{int}(k \pm | \phi_A, \phi_B)$, so both A and B signal to C. Moreover, the two conditions $\sum_{j}p^{int}(ijk \pm | \phi_A, \phi_B)= p^{int}(ik \pm | \phi_A, \phi_B)$ and $\sum_{i}p^{int}(ijk \pm| \phi_A, \phi_B)= p^{int}(jk \pm | \phi_A, \phi_B)$ mean respectively that B signals to A and C, and A signals to B and C. Therefore, we conclude that there is two-way signaling. Since the W-matrix is pure and the correlations can exhibit signaling in both directions A to B to C and B to A to C, we conclude that the process is causally nonseparable.

To summarise, in this paper we generalize the process matrix framework to continuous-variable quantum systems. This means that, as well as in finite dimensions, it is possible to describe the correlations between the measurement outcomes of two (or more) observers who can receive or send signals in absence of a global causally-ordered background. The correlations obtained are more general than those allowed by ordinary (causal) quantum mechanics. This generalization is suitable to devise new experiments using continuous-variable quantum systems, such as those considered in Gaussian quantum optics. Moreover, this work constitutes the first step towards the goal of formulating quantum fields on indefinite causal structures. As an example of application of this work, we implemented an infinite-dimensional version of the quantum switch exhibiting correlations stemming from quantum superposition of channels. 


\begin{acknowledgments}
	We thank Bernhard Baumgartner, Fabio Costa, Adrien Feix and Magdalena Zych for useful discussions. We acknowledge support from the European Commission project RAQUEL (No. 323970); the Austrian Science Fund (FWF) through the Special Research Program Foundations and Applications of Quantum Science (FoQuS), the doctoral programme CoQuS, and Individual Project (No. 24621).
\end{acknowledgments}

\end{document}